\documentclass[doublecol]{epl2}
\usepackage{amsmath}
\usepackage{amssymb}
\usepackage{amsthm}
\begin{document}
\def\PAu{Per$_2$[Au(mnt)$_2$]}
\def\PM{Per$_2$[$M$(mnt)$_2$]}
\title{Superconductivity close to the charge-density-wave  instability}
\author{H. Bakrim\inst{1} \and C. Bourbonnais\inst{1,2}}
\institute{                    
  \inst{1} Regroupement Qu\'eb\'ecois sur le Mat\'eriaux de Pointe, D\'epartement de physique, Universit\'e de Sherbrooke, Sherbrooke, Qu\'ebec, Canada, J1K-2R1\\
  \inst{2} Canadian Insitute for Advanced Research, Toronto, Ontario, Canada
}

\begin{abstract}
{We use  the weak coupling  renormalization group method  to examine  the interplay between charge-density-wave and s-wave superconducting orders  in  a quasi-one-dimensional  model of electrons interacting with acoustic phonons.  The relative stability of both types of order  is mapped out at  arbitrary nesting deviations and  Debye  phonon  frequency $\omega_D$.   We singled out  a    power law increase of the superconducting $T_c\sim \omega_D^{0.7}$  from a  quantum critical point  of  charge-density-wave order   triggered by nesting alterations. The results capture  the key features shown by the proximity between the two types of ordering  in the phase diagram of the  recently discovered Perylene based organic superconductor under pressure. The   impact of   Coulomb interaction  on the relative stability of the competing phases  is examined and  discussed in connection with the occurrence of  s-wave superconductivity in low dimensional charge-density-wave materials.}
\end{abstract}
\maketitle

\section{Introduction} The recent observation of superconductivity (SC) in the   Perylene based  organic conductor \PAu \cite{Graf09}, brings once again into focus  the possible role played by a  charge density-wave (CDW) instability in the  mechanism of onset of   SC  in   quasi-one-dimensional (quasi-1D) electron systems. This work makes use of the renormalization group (RG) technique to  analyze   the interplay between these two phases   in  systems of electrons     coupled to lattice vibrations. Besides their relevance for  materials showing a  CDW-SC proximity, the results also address the  issue of quantum criticality associated with interfering orders for models of electrons coupled  to bosonic excitations in low dimensions \cite{Demler01,Abanov03}.

\PAu\  is a member of  the two-chain charge transfer salts series \PM, where \hbox{$M=$} Pt, Pd, Au, $\ldots$. These organic salts are made of Perylene and Dithiolate flat molecular complexes that pile up as  segregated stacks   well described by a   quasi-1D  electronic structure \cite{Canadell04}. In normal pressure conditions, the Perylene chains undergo a  metal-insulator transition    due to the formation of a Peierls lattice distorted state driven by a  CDW  superstructure\cite{Henriques84}. For the \PAu\ compound, only the Perylene stacks are electronically active   in the  CDW transition, which  takes place at \hbox{$T_{\rm CDW}\simeq 12~$}K at ambient pressure  \cite{PeryleneSC2}. This is a relatively low temperature scale     likely to  be  vulnerable to nesting alterations  of the Fermi surface by the application of pressure. This is supported by  the suppression of the insulating state under 5~kbar of pressure, which turns out to be also critical to the onset of  SC     at $T_c\simeq 300~$mK \cite{Graf09}, hinting at a direct  part played  by  CDW correlations in the enhancement of  Cooper pairing.  The   sequence of states thus obtained is reminiscent of  the competition between CDW and SC orders found    in some quasi-1D transition-metal trichalcogenides materials \cite{Briggs80}.  The pattern is also akin to the quasi-1D Bechgaard salts series [(TMTSF)$_2X$], where  a  spin-density-wave (SDW) state is known to be followed  by superconductivity under pressure\cite{Jerome82,Vuletic02,DoironLeyraud09}.  In the latter   case the application of the RG method  to  electron models  with  momentum dependent repulsive interactions   has demonstrated  how   nesting deviations                                                                          can control the interference between density-wave and non s-wave Cooper pairings  and  reproduce the sequence of phases    displayed  by   the Bechgaard  salts under pressure\cite{Duprat01,PeryleneSC1,Bourbon09}.

  However, at variance with  SDW  systems  where  the direct Coulomb term  dominates  the scene of interactions, the electron-phonon coupling   plays  an essential   role in lattice distorted CDW systems. Electron-electron interactions induced  by the  exchange of acoustic phonons   are dynamically governed by the    Debye energy scale $\omega_D$, which is much smaller than the Fermi energy and often close to  the energy scale of  CDW order found in molecular conductors\cite{Jerome82}. This introduces retardation in interactions, which  besides   interchain hopping and nesting alterations,   modifies in a non trivial way the interfering  many-body processes that are linked to density-wave and Cooper pairings     in every order of perturbation theory. This  difficulty has been  well established  in the past,  requiring to go beyond the  habitual scheme of approximations such as  mean-field and RPA-like approaches that are   known to single out  one  pairing channel to the detriment of the other \cite{Barisic85a,Horovitz77,Caron84}.  In the one-dimensional case, a weak coupling solution to this problem  has been  found in the framework of the      RG  method \cite{Bakrim07,Tam07b}. Recent progress along these lines has shown that this approach   is well suited to simultaneously account for   both  pairing processes in the determination of ground states in  electron-phonon systems  at arbitrary phonon frequency.      
  
\begin{figure}
\onefigure[width=8.0cm]{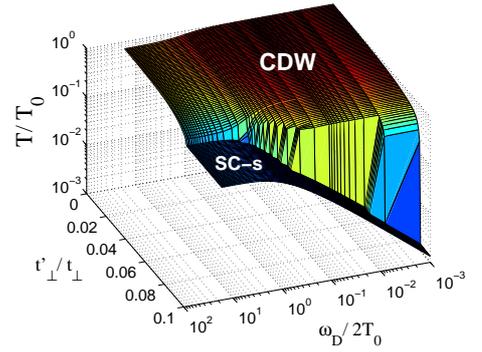}
\caption{Phase diagram of the quasi-1D electron-phonon model in the $(t_\perp',\omega_D)$ plane. Here  $T_0$ is the CDW ordering temperature of  the  adiabatic $\omega_D\to 0$  and perfect nesting $t_\perp'\to 0$ limit. }
\label{f1}
\end{figure}

In this paper  the  RG method is extended   to a electron-phonon model  in the quasi-1D case and at finite temperature. The temperature scales  $T_{{\rm{\tiny CDW}}}$ and $T_c$ for the instabilities of the metallic state against the formation of CDW and SC orders  are determined for arbitrary phonon frequency $\omega_D$ and nesting deviations    parametrized by the next-to-nearest neighbor interchain hopping $t_\perp'$. The main results of the present work are  outlined  in the phase diagram of Fig.~\ref{f1}.  For  small  $t_\perp'$, $T_{{\rm{\tiny CDW}}}$  weakens and   undergoes  a quantum-classical crossover as  $\omega_D$ is raised and goes beyond the adiabatic     scale $T_0$ for CDW ordering.
 When  nesting distortion attains    some threshold $t_\perp'^*$, $T_{{\rm{\tiny CDW}}}$ is critically reduced and at non zero  $\omega_D$,  an SC instability takes place in the s-wave channel only.  In the adiabatic  limit,   $T_{\rm{\tiny CDW}}$  defines a quantum critical point at $t_\perp'^*$, from which  an anomalous  power law increase of  $T_c$  with the phonon frequency takes place.  Along realsitic  pressure paths in the $(t_\perp',\omega_D) $ plane,  the model phase diagram follows  the  leading features  displayed by   Per$_2$[Au(mnt)$_2$]. The impact of the repulsive Coulomb interaction on the structure of the phase diagram is explored  alongside the predisposition of electron-phonon driven CDW systems  to show s-wave superconductivity.  

 \section{The model and the renormalization group equations}  We consider  a   non half-filled  two-dimensional electron  system consisting of $N_\perp $ chains of length $L$ with  the electron spectrum $E_p(\textbf{\textsl{k}})= v_F(pk-k_F) +\varepsilon_\perp(k_\perp)$, where $\varepsilon_\perp(k_\perp)=-2t_{\perp} \cos k_\perp - 2t'_\perp \cos 2 k_\perp $. Here $p=\pm$ refers to right and left moving electrons along the stacks, $v_F (k_F)$ to the parallel Fermi velocity (wave vector), and $t_\perp   $  to the interchain  hopping integral. In the quasi-1D case, we have $t_\perp \ll E_F= v_Fk_F$, where $E_F=E_0/2$ is the Fermi energy taken as  half   the band width. The next-to-nearest neighbor transverse hopping $t_\perp'$, which  describes nesting deviations, is   kept small compared to $t_\perp$. The following calculations are carried out  for the typical values $E_F= 15t_\perp =3000$~K.    In the framework of the Su-Schrieffer-Heeger model \cite{Su80}, the electrons are linearly coupled to parallel acoustic phonons. These modes being harmonic, this is equivalent  in the Matsubara formalism  of the  partition function  to consider a frequency dependent electron-electron interaction. In the g-ology picture, the  bare interaction reads

\begin{equation}
\label{ }
g_{i}(\bar{k}_1,\bar{k}_2;\bar{k}_3,\bar{k}_4) = { {g}^{\rm ph}_i\over 1+{(\omega_{n_1}-\omega_{n_4})^2/ \omega_D^2}},
\end{equation}
which for a  non half-filled band splits   as a  backward $(i=1)$ and forward  $(i=2)$ scattering amplitude between $p=+$ and $-$ moving carriers; here the momentum-frequency variables  $\bar{k}=(k_\perp,\omega_n)$ satisfy the conservation rule $\bar{k}_3 +\bar{k}_4=\bar{k}_1+\bar{k}_2$. In the following calculations we shall use the bare initial   amplitudes (normalized by $\pi v_F$) $ {g}^{\rm ph}_1 = -0.20$ for the backscattering part, while  $ { g}^{\rm ph}_2=0$ for  the forward scattering at vanishing momentum transfer -- the latter acquiring finite values following   the RG transformation. 

The RG transformation of the coupling constants results from the successive integration of electronic degrees of freedom in the outer energy shell $\pm  E_0(\ell)d\ell/2$ above and below the Fermi surface for all Matsubara frequencies. Here $E_0(\ell)= E_0e^{-\ell}$ is the scaled bandwidth at step $\ell\ge 0$. In the momentum-frequency   RG scheme  adopted here  at finite temperature, each  constant energy sheet from the Fermi surface is divided into  12 patches, each    defining a particular $k_\perp$ in momentum space, while a discrete set  of $N_\omega=14$ fermion Matsubara frequencies $\omega_n$ ($-7\le n\le 6$) is retained along the frequency axis. At finite temperature this represents a good compromise between exacting computing time and reproducing  the results  known for  either the non retarded case in  quasi-one dimension \cite{PeryleneSC1}  or the  electron-phonon problem in one dimension  \cite{Bakrim07}.

At the one-loop level, the  backward and forward  scattering amplitudes obey the flow equations 
\begin{align}
\label{RGg1}
& \partial_\ell g_{1}(\bar{k}_1,\bar{k}_2,\bar{k}_3,\bar{k}_4) =  {1\over 2\pi} \int dk_\perp I_P(k_\perp, \bar{q}_P) \cr
\times  \Big[ &\, \epsilon_P\langle g_1 (\bar{k}_1,\bar{k},\bar{k}_P,\bar{k}_4) g_1 (\bar{k}_P,\bar{k}_2,\bar{k}_3,\bar{k}) \rangle  \cr  
&+ \epsilon_{P,v} \langle g_2 (\bar{k}_1,\bar{k},\bar{k}_{4},\bar{k}_P) g_1 (\bar{k}_P,\bar{k}_2,\bar{k}_3,\bar{k}) \rangle \cr
&+ \epsilon_{P,v} \langle g_1 (\bar{k}_1,\bar{k},\bar{k}_P,\bar{k}_{4}) g_2 (\bar{k}_P,\bar{k}_2,\bar{k},\bar{k}_3) \rangle\Big]_{}    \cr
& + {1\over 2\pi} \int dk_\perp I_C(k_\perp, \bar{q}_C) \cr
\times \Big[ &  \epsilon_{C} \langle g_1 (\bar{k}_1,\bar{k}_2,\bar{k},\bar{k}_C) g_2 (\bar{k},\bar{k}_C,\bar{k}_4,\bar{k}_3) \rangle  \cr
    &+   \epsilon_{C}  \langle g_2 (\bar{k}_1,\bar{k}_2,\bar{k}_C,\bar{k}) g_1 (\bar{k},\bar{k}_C,\bar{k}_3,\bar{k}_4) \rangle\Big] 
    \end{align}
    and 
   \begin{align}
    \label{RGg2}
  & \partial_\ell g_{2}(\bar{k}_1,\bar{k}_2,\bar{k}_3,\bar{k}_4) =   {1\over 2\pi} \int dk_\perp  I_P(k_\perp, \bar{q}_P') \cr
\times  & \quad \epsilon_{P,l} \langle g_2 (\bar{k}_1,\bar{k},\bar{k}_3,\bar{k}'_P) g_2 (\bar{k}'_P,\bar{k}_2,\bar{k},\bar{k}_4) \rangle  \cr 
& +  {1\over 2\pi} \int dk_\perp  I_C(k_\perp, \bar{q}_C) \cr
\times \Big[& \,\epsilon_{C}  \langle g_1 (\bar{k}_1,\bar{k}_2,\bar{k},\bar{k}_C) g_1 (\bar{k},\bar{k}_C,\bar{k}_4,\bar{k}_3) \rangle \cr
       &\quad  + \epsilon_{C} \langle g_2 (\bar{k}_1,\bar{k}_2,\bar{k}_C,\bar{k}) g_2 (\bar{k},\bar{k}_C,\bar{k}_3,\bar{k}_4) \rangle\Big].
\end{align}
 These consist of  closed loop ($\epsilon_P=-2$), vertex corrections  ($\epsilon_{P,v}=1$) and ladder ($\epsilon_{P,l}=1$) diagrams of the 2$k_F$ electron-hole (Peierls) pairing,   which combine with  the  ladder diagrams ($\epsilon_C=-1$) of the electron-electron (Cooper) pairing. Here  $\bar{k}_P= \bar{k} + \bar{q}_P$   ,$\bar{k}'_P= \bar{k} + \bar{q}_P'$ and  $\bar{k}_C= -\bar{k} +  \bar{q}_C$, where  $ \bar{q}_{P,C}= (q_{\perp P,C}, \omega_{P,C})$   corresponds to the Peierls $ \bar{q}_P=\bar{k}_1 -\bar{k}_4$,  $ \bar{q}'_P=\bar{k}_1 -\bar{k}_3$ and Cooper $\bar{q}_C=\bar{k}_2 +\bar{k}_1$ variables.  In the above equations, each diagram contains a frequency convolution of the form $\sum_{\omega_n} g_i g_j{\cal L}_{C,P}$, which    has been decoupled as  $\langle g_ig_j\rangle   \sum_{\omega_n}{\cal L}_{C,P}$ at finite temperature. Here $\langle \cdots \rangle  ={N_\omega}^{-1}\sum_{n} \cdots$, stands as an average  of the couplings over the internal loop frequency variable, whereas the $\ell$ derivative of the     Cooper and Peierls loops  $I_{P,C}= \sum^{+\infty}_{n=-\infty}{\cal L}_{P,C}$ is evaluated exactly to give \begin{align}
\label{IPC}
   I_{P,C}&(k_\perp,\bar{q}_{P,C})=  \sum_{\nu=\pm 1} \Theta[|E_0(\ell)/2 + \nu A_{P,C}|- E_0(\ell)/2]   \cr
    &\times {1\over 4} \left[\tanh {E_0(\ell) + 2 \nu A_{P,C}\over4T} + \tanh {E_0(\ell)\over 4T}\right]\cr
    &\ \ \  \ \ \times \frac{(E_0(\ell) + \nu A_{P,C})E_0(\ell)}{(E_0(\ell) + \nu A_{P,C})^{2}+ \omega^{2}_{P,C}},  
\end{align}where $A_P= -\varepsilon(k_\perp) -\varepsilon(k_\perp+q_{\perp P})$, $A_C=   -\varepsilon(k_\perp) + \varepsilon(k_\perp+ q_{\perp C})$, and $\Theta(x)$ is the step function ($\Theta(0)\equiv {1\over2}$).

\begin{figure}
\onefigure[width=8.7cm]{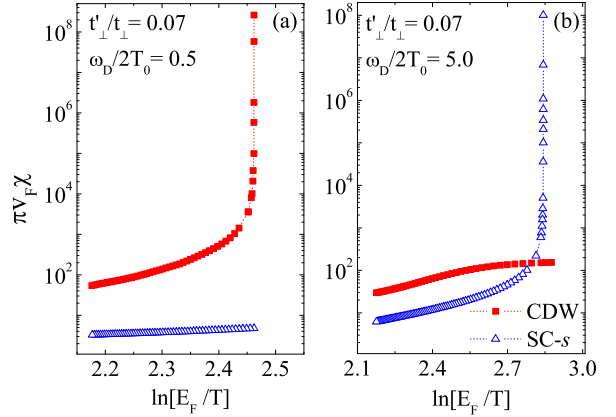}
\caption{Typical low temperature dependence of the charge-density-wave ($\chi_{\rm CDW}$: squares) and s-wave superconducting ($\chi_{\rm SC}$: triangles) normalized susceptibilities  in  the a)  CDW and b) SC ordering sectors of Fig.~\ref{f4}. }
\label{f2}
\end{figure}

 A singularity   can occur   in either the Peierls  or the Cooper scattering channel,   which is indicative of an  instability of the metallic state towards long-range order at the temperature $T_\mu$.  The nature of the ordering state is provided by the singularity  of the  static  and normalized  response function  $\pi v_F\chi_\mu( {\bf q}^0_\mu) = (2\pi)^{-1} \iint   dk_\perp d\ell\,  \langle z^2_\mu(\bar{k})\rangle   I_{P,C}(k_\perp,{q}^0_{\perp,\mu}) $ at the wave vector ${\bf q}^0_{\rm CDW}=(2k_F,\pi)$ for $\mu={\rm CDW}$ and ${\bf q}^0_{\rm SC}=0$ for $\mu ={\rm SC}$. The     response vertex parts $z_\mu$ are governed by the  equations 
  \begin{align}
\label{RGCDW}
& \partial_\ell  z_{\rm CDW}(\bar{k}+ \bar{q}_P^0) =   {1\over 2\pi} \int dk'_\perp I_P(k_\perp',\bar{q}^0_P) z_{\rm CDW}(\bar{k}' + \bar{q}_P^0)\cr 
&\! \! \times \! \big\langle  [ \epsilon_Pg_1(\bar{k}'\!+\bar{q}_P^0,\bar{k}, \bar{k}+\bar{q}_P^0,\bar{k}') \! \cr
& \hskip 2 truecm  + \epsilon_{P,v}g_2(\bar{k}'\!+\bar{q}_P^0,\bar{k}, \bar{k}'\!,\bar{k}+\bar{q}_P^0)]\big\rangle,
\end{align}
for the  $\mu=$CDW response ($\bar{q}_P^0=(\pi,0)$) and
 \begin{align}
\label{RGSC } 
& \partial_\ell  z_{\rm SC}(-\bar{k}+\bar{q}_C^0) =   {1\over 2\pi} \int dk'_\perp I_C(k_\perp',\bar{q}^0_C) z_{\rm SC}(-\bar{k}'+\bar{q}_C^0)\cr 
&\! \! \times \! \big\langle \epsilon_C [ g_1(-\bar{k}'\!+\bar{q}_C^0,\bar{k}', -\bar{k}+\bar{q}_C^0,\bar{k})   \cr
&      \hskip 2 truecm + g_2(-\bar{k}'+\bar{q}_C^0,\bar{k}',\bar{k},-\bar{k}+\bar{q}_C^0)]\big\rangle,
\end{align} 
for the static (s-wave) $\mu=$SC response  ($\bar{q}_C^0=0$). For the whole range of parameters covered by  the present   model, the  finite temperature singularities  only   occur for either the  CDW or  s-wave  SC susceptibilities  (Fig.~\ref{f2}). 
 \section{Results} 
 Let us first consider the instability of the metallic state as one moves along the phonon frequency axis at fixed   $t_\perp'$ (Fig.~\ref{f1}). At perfect nesting    $t_\perp'=0$, the adiabatic limit   $\omega_D \to 0$ is characterized by a singularity    signaling the occurrence of a Peierls instability at the temperature denoted   $T_0$ ($\simeq 20$K for the parameters chosen here). In this limit, only close loops  contribute to the flow of Eqs.~(\ref{RGg1}-\ref{RGg2}) and (\ref{RGCDW}),  a limit  equivalent  to  the molecular field  analysis of the  Peierls  instability of the metallic state. By increasing  $\omega_D$, both the vertex and ladder diagrams are progressively unlocked and  begin to mix and interfere with closed loops. In the pertinent temperature range $T\ll t_\perp$ where all the  instabilities  take place, the transverse electronic motion is coherent. As a function of energy, the interference   is then  maximum in the one-dimensional part  of the flow where $E_0(\ell)/2 > t_\perp$, whereas  for  $E_0(\ell)/2 < t_\perp$   interchain hopping begins to be coherent and the interference    becomes non uniform in momentum space and generates a $k_\perp$  dependence of the coupling constants \cite{Duprat01,PeryleneSC1}. 
\begin{figure}
\onefigure[width=7.0cm]{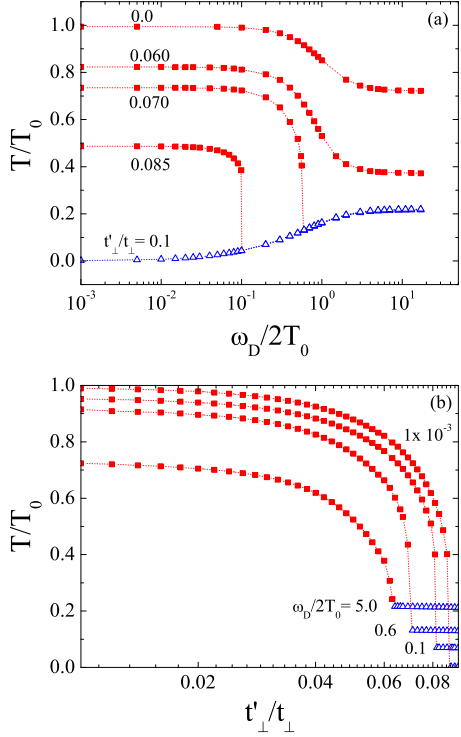}
\caption{ (a)  Normalized $T_{\rm CDW}$ (square)  and s-wave $T_c$ (triangle) {\it vs} the phonon frequency ratio $\omega_D/2T_0$ at different   $t_\perp'/t_\perp$. (b) Normalized $T_{\rm CDW}$ (square)  and s-wave $T_c$ (triangle)  {\it vs} the nesting deviation parameter $t_\perp'/t_\perp$ for different frequency ratios $\omega_D/2T_0$.  }
\label{f3}
\end{figure}
As a result, $T_{\rm CDW}$ diminishes  with   increasing $\omega_D$. However, when the frequency reaches the classical Peierls scale $\omega^*_D(t_\perp'=0) \sim 2 T_0$,  the decrease is more rapid and $T_{\rm CDW}$ undergoes  a crossover toward a  non adiabatic CDW  regime where all diagrams of both pairing channels contribute   and ultimately  level  off  the reduction of  $T_{\rm CDW}$ (Fig.~\ref{f3}-a) -- a crossover analogous to  the one found in the purely 1D case at $t_\perp=0$\cite{Bakrim07,Caron84}.  

A  finite  amplitude of the  anti-nesting  term  $t_\perp'$ modifies $A_P$ in (\ref{IPC}), which reduces all the Peierls diagrams but   leaves  those of the  Cooper channel unchanged.    Thus $T_{\rm CDW}$ first  gets	 smaller (Fig.~\ref{f3}-a),   then drops rapidly  as the phonon frequency  extends across   $\omega_D^*$, which is also smaller,   to  finally attain  the non adiabatic limit (Fig.~\ref{f3}-a). This reduction carries on  until $t_\perp'$ reaches in its turn  a critical value  where  $\omega_D^*$ signals a crossover toward  a different instability of the metallic state where the Cooper pairing processes are prevailing  and   the instability is against s-wave SC     at $T_c$. According to Fig.~\ref{f3}-a, the SC critical temperature increases with   $\omega_D$  and  finally reaches a plateau above   $\omega_D\sim 2T_0$,  a scale apparently still tied to the adiabatic  CDW  limit.  The   s-wave Cooper pairing attraction taking place along the chains and is  strongly enhanced by CDW fluctuations  (Fig.~\ref{f1}-b).  The     $T_c$ values thus achieved are markedly enhanced with respect to   the BCS  limit obtained when   all the Peierls loops in (\ref{RGg1}-\ref{RGg2}) are put to zero.  

\begin{figure}
\onefigure[width=8.0cm]{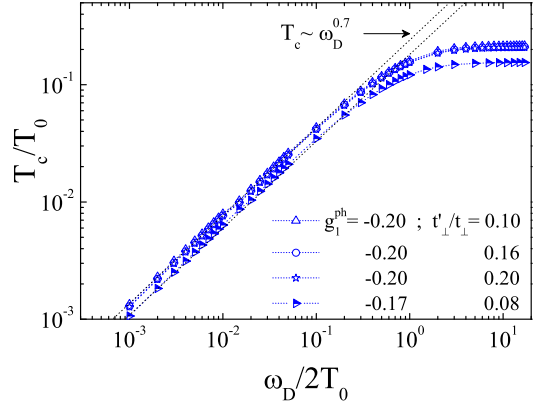}
\caption{Phonon frequency dependence of the superconducting  ordering temperature $T_c$  at different   $t_\perp'> t_\perp'^*$ and electron-phonon coupling $g_1^{\rm ph}$. The dashed lines correspond to the  power  law $T_c\sim \omega_D^{0.7}$.  }
\label{f4}
\end{figure}

When  in the adiabatic  limit,  $t_\perp'$ is further increased  up the critical value $t_\perp'^*\approx 0.9T_0$, nesting alterations are sufficiently large to suppress the singularity of the Peierls channel, bringing $T_{\rm CDW}$ down to zero (top curve, Fig.~\ref{f3}-b). Since at    $\omega_D\to 0$, all the ladder diagrams are vanishingly small, the point $t_\perp'^*$ defines a quantum critical point for CDW ordering. Moving now away from this point,  along the frequency axis, the SC instability  shows up  as a result of the ladder diagrams of the Cooper channel that  progressively unfold.  However, at variance with the single channel BCS approximation, $T_c$  follows a power law increase $T_c\sim \omega_D^\eta$, with an exponent $\eta \simeq 0.70$    smaller than unity (Fig.~\ref{f4})  -- the BCS   value  $\eta\simeq1$ being recovered when all the Peierls loops are turned off.  

 Being independent of $t_\perp'>t_\perp'^*$ and  for a sizable range   $g^{\rm ph}_1$ in weak coupling, the exponent $\eta$ shows no noticeable trace of  non universality  (Fig.~\ref{f4}). The non BCS increase of $T_c$ is a direct consequence of  the influence of CDW correlations on Cooper pairing and  which takes place at all energy scales. As a matter of fact, when $\omega_D$ increases, the Cooper diagrams grow  in importance on one hand, but CDW correlations and then the pairing interaction is reduced on the other. It is the  combination of both effects that leads to an exponent $\eta$ smaller than unity.  This contrasts with the BCS case where only the former effect is present, while  the  coupling is considered essentially fixed and  attractive only below the sharp cutoff $E_0(\ell)/2 =\omega_D$ \cite{PeryleneSC3}. 
 
 The `critical line' $T_c\sim \omega_D^\eta$ corresponds to a quantum-classical transition  between the metallic and the SC states. The Debye frequency  can thus be put in the category of a symmetry breaking parameter that drives the transition at a fixed  electron-phonon coupling. In the standard terminology  \cite{Sachdev99},    $\eta$ is  the crossover exponent $\phi=z\nu$ of the transition expressed in terms of the product of the dynamical and coherence length exponents. Assuming Lorentz invariance of the model,   this forces  $z=1$, which would   imply  an anomalous dimension for the coherence length exponent, namely  $\nu=\eta$. 
Interestingly, if one looks at the range of phonon frequency over which the power law for $T_c$ takes place, one realized that it  is confined to low frequency. According to Figs.~\ref{f2}-a and \ref{f3}, $T_c$ indeed levels off when $\omega_D$ exceeds the  Peierls scale $2T_0$ for non adiabaticity,  stressing once again the non BCS character of the transition.  

To  complete   the analysis  of the transition profile  as a function of nesting deviations, one observes from  Fig.~\ref{f3}-b that at finite $\omega_D$,   the weakening of $T_{\rm CDW}$  by ladder diagrams and vertex corrections is correlated to  a reduction of the   threshold value $t_\perp'^*$ for the   onset of superconductivity.  The decrease of   $T_c$  with $t_\perp'>  t_\perp'^*$ is found in Fig.~\ref{f3}-b to be relatively slow for any  finite $\omega_D$. In effect,   $t_\perp'$  is  more effective  as a low energy scale to cut the CDW singularity than to  CDW correlations responsible of the major part of the attractive pairing (Fig.~\ref{f2}-b).   

 The above  one-loop RG results    for the present model
     are  summed up in the global phase diagram of Figure~\ref{f1} for arbitrary nesting deviations and phonon frequency.

\section{Connection with experiments}
The above results can apply to low dimensional CDW systems where  the electron-phonon interaction is the prevalent mechanism for ordering. This is distinctly possible in CDW compounds  like Per$_2$[Au(mnt)$_2$] for which the flat Perylene molecular unit   is rather large in size and polarizable. The resulting Coulomb interaction is thus expected to be   small as corroborated by the weak enhancement of the electron spin susceptibility and nuclear spin-lattice relaxation rate  reported for  this material \cite{Almeida97,Bourbon91b}.  The  electron-phonon model considered above can then in a first approximation  be applied    to the CDW-SC sequence displayed by the compound   under pressure. Taking  into consideration the pronounced quasi-1D anisotropy of this material \cite{Canadell04},  the $T_{\rm CDW}$($\simeq 12$K \cite{Graf09,PeryleneSC2})  observed  in normal pressure conditions can be considered not   too far below    the optimal scale $T_0$ calculated  at perfect nesting. The Debye frequency  for the $2k_F$ acoustic phonons of the quarter-filled Perylene  stacks, though  not known  with accuracy, can be  at least be taken as few   dozens of degrees. This fairly places the compound with a   frequency ratio \hbox{$\omega_D/ 2T_0> 1$},   and according to Fig.~\ref{f1}, with favorable  conditions for    superconductivity under pressure.    As pressure scales up  both band parameters and phonon frequency \cite{PeryleneSC4},  the system is likely to move from the CDW region ($t'_\perp< t_\perp'^*,\omega_D/2T_0> 1)$ toward    SC   where ($t'_\perp\gtrsim t_\perp'^*,\omega_D/2T_0> 1)$ in the $(t_\perp', \omega_D)$  plane of Fig.~\ref{f1}, a path congruent with  the  results of  Graf {\it et al.}  \cite{Graf09}.

 The  pairing attraction of the model, albeit boosted by  CDW correlations, takes place along the chains and   is responsible for the s-wave character of superconductivity. While a  s-wave order parameter is well known to sustain the presence of non magnetic impurities; on the contrary,  it  is sensitive to Coulomb interaction that    is finite in practice and acts as a pair breaking effect. The impact  the Coulomb term has  on the above results can be  readily examined by modifying the  initial conditions  of the flow equations (\ref{RGg1}) and (\ref{RGg2}). In the framework of the extended Hubbard model, this amounts  to add the constant terms $g_1=U$ and $g_2=U +2V$ to the backward and forward scattering amplitudes at quarter-filling. Here  $U>0$ and $V>0$ are the on-site and first nearest-neighbor repulsive {\it intra}-chain interaction  parameters of the extended-Hubbard  model, here normalized by $\pi v_F$. 
\begin{figure}
\onefigure[width=8.7cm]{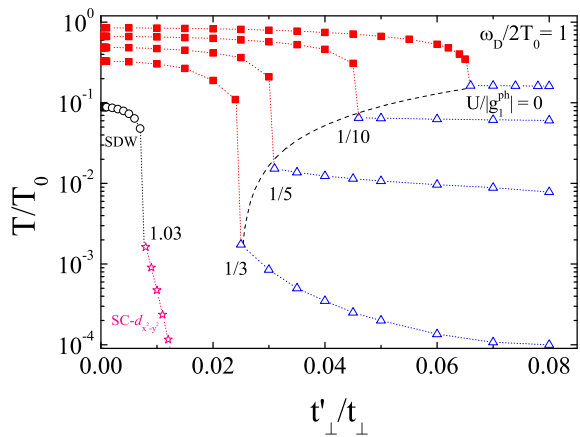}
\caption{Variation of the transition temperature ($T_{\rm CDW}$: squares; $T_c$: triangles)  as a function of the antinesting parameter $t_\perp'$ for different Coulomb interaction amplitude $U/|g_1^{\rm ph}|$, normalized by $|g_1^{\rm ph}| $, along $U=2V$, and  for $\omega_D/2T_0=1$. The dashed line gives   the reduction of the maximum $T_c$ with $U$. The circles (stars) refer to  the   SDW (d-wave SC) instability emerging  for  $U/|g_1^{\rm ph}|\gtrsim 1$. }
\label{f5}
\end{figure}

 From the foregoing analysis   and for suitable  conditions for superconductivity  at $\omega_D/2T_0 \sim 1$, the one-loop RG  solution for the critical temperature along the line $U=2V$ is given in Fig.~\ref{f5}  as a function of the antinesting parameter $t_\perp'$.  While weak intra-chain Coulomb interaction  reduces  slightly  $T_{\rm CDW}$ (essentially due to the  repulsive backscattering component $g_1$),  its detrimental impact on superconductivity is particularly pronounced for relatively weak repulsive interactions. The  maximum $T_c$  at $t_\perp'^*$ drops   by  an order of magnitude at $ U \sim   |g^{\rm ph}_1|/5$. However, it worth noticing that    such a range of    $U$  is sufficient to   reduce    the ratio $T_c/T_{\rm CDW}|_{\rm max}$     to    values  comparable   with the one found  experimentally in Per$_2$[Au(mnt)$_2$] \cite{Graf09} and in the inorganic trichalcogenide compound   NbSe$_3$ \cite{Briggs80}.  
 
      By reinforcing repulsive interactions, $T_c$ will therefore not be very long to  become vanishingly small and potentially undetectable in practice, in spite of a  sizable  $T_{\rm CDW}$ at low pressure. This may supply  some  insight as to why   quarter-filled molecular compounds  like TMTSF-DMTCNQ \cite{Andrieux79a}, known as  
      a correlated    CDW system \cite{Pouget81,Jerome82},  failed to show any sign of superconductivity following the suppression of  its lattice distorted state  under pressure.  Such an interpretation may also adhere to  the absence of  superconductivity in the  phase diagram of correlated quasi-1D compounds like  [EDT-TTF-CONMe$_2$]X \cite{Auban09}, and  (DI-DCNQI)$_2$X \cite{Itou05}  at high pressure.  From this angle, the chance   for a  correlated organic metal like  TTF-TCNQ \cite{Jerome82}, whose CDW order  is expected to vanish around 90 kbar\cite{Yasuzuka07},    to show the presence of superconductivity  is reduced at the very least in the s-wave channel. 
       
     We finally examine in Fig. \ref{f5}  the impact of further  increasing the Coulomb term $U$ on the phase diagram, namely  beyond the amplitude of the  phonon induced interaction.  A qualitative change then occurs in the nature and the sequence of ground states. The CDW  gives way to a SDW instability at low  $t_\perp'$, which  in its turn  yields non conventional d-wave superconductivity at  $t_\perp'^*$ and above. It is where the results of the present work  connect  to the sequence of instabilities found in  the purely repulsive case \cite{Duprat01,PeryleneSC1,Bourbon09}, which is known to apply in systems like the Bechgaard salts \cite{Jerome82,Vuletic02,DoironLeyraud09}.  
       
  \section{Conclusion} In conclusion the instability of the metallic state against charge-density-wave and s-wave superconducting orders  in  quasi-1D systems can be analyzed  by extending the  RG approach to electrons coupled  to phonons of arbitrary frequency. The results show that  both instabilities influence one another  and form  a sequence of  ordered states that captures the key traits of  low dimensional charge density-wave materials exhibiting superconductivity  under pressure.


\acknowledgments
 C. B thanks the National Science and Engineering Research Council  of Canada (NSERC), the R\'eseau Qu\'ebcois des Mat\'eriaux de Pointe (RQMP) and  the {\it Quantum materials} program of Canadian Institute of Advanced Research (CIFAR) for financial support. Computational resources were provided by the R\'eseau qu\'eb\'ecois de calcul de haute performance (RQCHP) and Compute Canada.


 \bibliography{/Users/cbourbon/Dossiers/articles/Bibliographie/articlesII.bib}

\begin{thebibliography}{10}

\bibitem{Graf09}
D. Graf {\it et~al.}, Eur. Phys. Lett. {\bf 85},  27009  (2009).

\bibitem{Demler01}
E. Demler, S. Sachdev, and Y. Zhang, Phys. Rev. Lett. {\bf 87},  06702  (2001).

\bibitem{Abanov03}
A. Abanov, A.~V. Chubukov, and J. Schmalian, Adv. Phys. {\bf 52},  119  (2003).

\bibitem{Canadell04}
E. Canadell, M. Almeida, and J. Brooks, Eur. Phys. J. B {\bf 42},  453  (2004).

\bibitem{Henriques84}
R.~T. Henriques, L. Alcacer, J.~P. Pouget, and D. Jerome, J. Phys. C {\bf 17},
  5197  (1984).

\bibitem{PeryleneSC2}
G. Bonfait {\it et al.,} Solid State Commun. {\bf 80}, 391 (1991); G. Bonfait,
  M. J. Matos, R. T. Henriques, and M. Almeida, Physica B {\bf 211}, 297
  (1995).

\bibitem{Briggs80}
A. Briggs {\it et~al.}, J. Phys. C {\bf 13},  2117  (1980).

\bibitem{Jerome82}
D. J{\'e}rome and H.~J. Schulz, Adv. Phys. {\bf 31},  299  (1982).

\bibitem{Vuletic02}
T. Vuletic {\it et~al.}, Eur. Phys. J. B {\bf 25},  319  (2002).

\bibitem{DoironLeyraud09}
N. Doiron-Leyraud {\it et~al.}, Phys. Rev. B {\bf 80},  214531  (2009).

\bibitem{Duprat01}
R. Duprat and C. Bourbonnais, Eur. Phys. J. B {\bf 21},  219  (2001).

\bibitem{PeryleneSC1}
J. C. Nickel and R. Duprat and C. Bourbonnais and N. Dupuis, Phys. Rev. Lett.
  {\bf 95}, 247001 (2005); Phys. Rev. B {\bf 73}, 165126 (2006).

\bibitem{Bourbon09}
C. Bourbonnais and A. Sedeki, Phys. Rev. B {\bf 80},  085105  (2009).

\bibitem{Barisic85a}
S. Barisic,  in {\em Electronic Properties of Inorganic Quasi-One-Dimensional
  Compounds}, edited by P. Monceau (D. Reidel, Dordrecht, Holland, 1985),
  Vol.~Physics and Chemistry of materials with low-dimensional structures,
  Series B, Quasi-one-dimensional Materials, Part I, p.\ 1.

\bibitem{Horovitz77}
B. Horovitz, Phys. Rev. B {\bf 16},  3943  (1977).

\bibitem{Caron84}
L.~G. Caron and C. Bourbonnais, Phys. Rev. B {\bf 29},  4230  (1984).

\bibitem{Bakrim07}
H. Bakrim and C. Bourbonnais, Phys. Rev. B {\bf 76},  195115  (2007).

\bibitem{Tam07b}
K.-M. Tam, S.-W. Tsai, D.~K. Campbell, and A.~H.~C. Neto, Phys. Rev. B {\bf
  75},  161103(R)  (2007).

\bibitem{Su80}
W.~P. Su, J.~R. Schrieffer, and A.~J. Heeger, Phys. Rev. B {\bf 22},  2099
  (1980), see also S. Barisic, Phys. Rev. B {\bf 5}, 932 (1972).

\bibitem{PeryleneSC3}
It is worth stressing that in the framework of the BCS-Elisahberg theory of
  superconductivity, the $\omega_D$ dependent reduction of the Coulomb
  pseudo-potential $\mu^*$ by the scattering in the electron-electron channel
  is well known to yield a power law isotope effect for $T_c\propto
  \omega_D^\alpha$, with an exponent $\alpha$ that differs from unity. See for
  example, J. P. Carbotte, Rev. Mod. Phys. {\bf 62}, 1027 (1990).

\bibitem{Sachdev99}
S. Sachdev, {\em Quantum Phase Transitions} (Cambridge Univ. Press, Cambridge,
  U.K., 1999).

\bibitem{Almeida97}
M. Almeida and R.~T. Henriques,  in {\em Handbook of Organic Conductive
  Molecules and Polymers}, edited by H. Nalwa (Wyley, New York, 1997), p.\ 87.

\bibitem{Bourbon91b}
C. Bourbonnais {\it et~al.}, Phys. Rev. B {\bf 44},  641  (1991).

\bibitem{PeryleneSC4}
Strickly speaking pressure also reduces the initial backscattering amplitude of
  $|g_1^{\rm ph}|\propto\kappa^{-1}$, mainly through the hardening of the
  phonon spring constant $\kappa$, an effect that has not been included here
  and which would lead to an additional but smooth reduction of both $T_{\rm
  CDW} $ and $T_c$ under pressure.

\bibitem{Andrieux79a}
A. Andrieux {\it et~al.}, J. Physique. Paris {\bf 40},  1199  (1979).

\bibitem{Pouget81}
J.~P. Pouget, Chemica Scripta {\bf 55},  85  (1981).

\bibitem{Auban09}
P. Auban-Senzier {\it et~al.}, Phys. Rev. Lett. {\bf 102},  257001  (2009).

\bibitem{Itou05}
T. Itou {\it et~al.}, Phys. Rev. B {\bf 72},  113109  (2005).

\bibitem{Yasuzuka07}
S. Yasuzuka, K. Murata, T. Arimoto, and R. Kato, J. Phys. Soc. Jpn {\bf 76},
  033701  (2007).

\end{thebibliography}
\bibliographystyle{prsty}
\end{document}